\documentclass[letter]{aa} 

\usepackage{graphicx}
\usepackage{txfonts}
\usepackage{hyperref}
\hypersetup{linkcolor=blue,citecolor=blue,colorlinks=true,pdfborderstyle={}}
\def\cgskev{erg~cm$^{-2}$~s$^{-1}$~keV$^{-1}$}

\begin{document} 

   \title{The neutrino background from non-jetted active galactic nuclei}


   \author{P. Padovani 
          \inst{1}
     \and 
           R. Gilli
          \inst{2}
          \and 
           E. Resconi
          \inst{3}
          \and 
           C. Bellenghi
          \inst{3}
          \and 
           F. Henningsen
           \inst{4}
          }

   \institute{European Southern Observatory,
              Karl-Schwarzschild-Stra{\ss}e 2, D-85748 Garching bei M{\"u}nchen, Germany
              \email{ppadovan@eso.org}
               \and 
               INAF -- Osservatorio di Astrofisica e Scienza dello Spazio, via Piero Gobetti 93/3, 40129 Bologna, Italy
              \and
                Technische Universit\"at M\"unchen, TUM School of Natural Sciences, Physics Department, 
James-Frank-Stra{\ss}e 1, D-85748 Garching bei M{\"u}nchen, Germany
            \and
            Simon Fraser University, Department of Physics, 8888 University Dr, Burnaby, Canada V5A 1S6
             }

   \date{Received 19 March 2024; accepted 5 April 2024}

 
  \abstract
   {}
  {We calculate the contribution to the neutrino background from the non-jetted active galactic nuclei (AGN) population following the recent IceCube association of TeV neutrinos with NGC 1068.}
   {We exploit our robust knowledge of the AGN X-ray luminosity function and evolution and convert it to the neutrino band by using NGC 1068 as a benchmark and a theoretically motivated neutrino spectrum.}
   {The resulting neutrino background up to redshift 5 does not
   violate either the IceCube diffuse flux or the upper bounds for
   non-jetted AGN, although barely so. This is consistent with a scenario where the latter class makes a substantial contribution mostly below 1 PeV, while jetted AGN, i.e. blazars, dominate above this energy, in intriguing agreement with the dip in the neutrino data at 
$\sim 300\,$TeV. More and better IceCube data on Seyfert galaxies will allow us to constrain the fraction of neutrino emitters among non-jetted AGN.}
   {}


   \maketitle
%

\section{Introduction}\label{sec:intro}

IceCube has recently detected an excess of TeV neutrinos with a 4.2\,$\sigma$ significance
from the direction of 
NGC 1068, the prototype Seyfert II galaxy \citep{Abbasi_2022}. 
This event was somewhat unexpected. Non-jetted AGN\footnote{We follow \cite{Padovani_2017NatAs} and define as jetted AGN
with strong relativistic jets, which is not the case for NGC 1068, as discussed in \cite{Padovani_2024}.} such as NGC 1068, 
in fact, are usually characterised by thermal emission, in contrast to jetted AGN, known for 
their predominantly non-thermal radiation \citep[e.g.][]{Padovani_2017}. Only the latter class 
was deemed to have the capability to accelerate protons to the energies necessary for neutrino 
production, as discussed, for example, in the review by \cite{Giommmi_2021} (and references therein; but see \citealt{berezinsky1977,eichler1979,silberberg1979}).

Being so close, NGC 1068 can be spatially resolved into a number 
of components, all possibly relevant to neutrino production \citep{Padovani_2024}. 
These include: 1. a starburst 
region in the spiral arms of its host galaxy; 2. a $\lesssim$ kpc jet; 3. a sub-kpc molecular
outflow; 4. and the supermassive black hole (SMBH) vicinity. 
By first using simple order-of-magnitude arguments and then applying 
specific theoretical models, \cite{Padovani_2024} (and references therein) have come to the conclusion 
that only the region close to the accretion disc around the SMBH, most likely 
the X-ray emitting corona, fulfils the conditions to have both the right density of photons 
needed to provide the targets for protons to sustain neutrino production and to 
absorb the expected but unobserved $\gamma$ rays. 

How unique is NGC 1068? At present the extent to which the process that produces 
neutrinos in NGC 1068 generalise to the wider AGN population is unknown. 
Assuming a tight connection between neutrino emission and
the plasma in AGN coronae, theoretical studies of Seyfert galaxies that host 
X-ray bright AGN have been carried out to single out possible new neutrino sources \cite[e.g.][]{Kheirandish_2021}. 

The purpose of this Letter is to approach this question from an observational and population-wide perspective. We leverage the well-established knowledge of X-ray luminosity functions (XLF) and evolution of AGN, along with the constraints provided by the cosmic X-ray background (XRB; \citealt{Gilli_2007}). We then translate these findings to the neutrino band by normalizing them to NGC 1068. In short, our aim is to utilize the limited neutrino data, both on the observational and modelling side, to investigate the implications of extrapolating from a single non-jetted AGN to the entire population.
We assume a distance to this source
of 10.1 Mpc, following \cite{Padovani_2024}. 


\section{Population synthesis of the X-ray background}
The spectrum of the XRB records the integrated emission of AGN across all cosmic times \citep{Setti_1989}. Several papers \citep[e.g.][]{Comastri_1995,Treister_2006,Akylas_2012,Ueda_2014,Ananna_2019} demonstrated
that the peak around $20-30$ keV in the XRB spectrum cannot be reproduced by the emission of unobscured and moderately-obscured AGN alone, but that a large
population of heavily obscured, Compton-thick AGN\footnote{These are defined by a column density along our line-of-sight $N_{\rm H}>1/\sigma_T\sim 10^{24}$cm$^{-2}$, where $\sigma_T$ is the Thomson cross-section.} is required. To date, population synthesis models of the XRB provide the best inference about the overall abundance of accreting SMBHs. We adopt here the model of \citet{Gilli_2007} which, besides accurately fitting the XRB spectrum, is in very good agreement with the most recent estimates of the AGN XLFs (\citealt{Ueda_2014}) and number counts in different redshift intervals and X-ray bands \citep{Luo_2017,Nanni_2020, Marchesi_2020}. Starting from the 
$0.5-2$ keV XLF of unobscured AGN described in \citet{HMS_2005}, and assuming a distribution of AGN obscuring column densities in agreement with observational constraints \citep{Risaliti_1999,Tozzi_2006}, \citet{Gilli_2007} found that moderately obscured, Compton-thin AGN ($N_{\rm H}=10^{22-24}$cm$^{-2}$) need to outnumber unobscured AGN (by a factor decreasing from 4 to 1 with increasing luminosity) to reproduce the measured $2-10$ keV AGN XLF \citep{Ueda_2003, Lafranca_2005}. Furthermore, a luminosity-dependent space density of Compton-thick AGN equal to that of moderately obscured Compton-thin AGN was required to fit the XRB spectral peak (assuming all populations follow the same cosmological evolution). We remark that the AGN XLF used in the \citet{Gilli_2007} are largely ($\gtrsim90\%$) dominated by non-jetted AGN. As discussed below, our computation will specifically refer to the contribution to the neutrino background from the non-jetted AGN population only.

\section{Predicting the neutrino background}\label{sec:predict}

To compute the expected neutrino background using CXB population synthesis models we first extended the AGN X-ray spectra considered by \citet{Gilli_2007} to the neutrino domain.

We first assumed that the $10^{-4}-10^3$ TeV neutrino spectrum of each non-jetted AGN follows the ``minimal $pp$ scenario'' model of \cite{Murase_2022} (as shown in their Fig.~3, left and middle panels), 
which provides an X-ray corona-based theoretical fit to the NGC 1068 IceCube data.
Such a spectrum peaks at $E\sim 1$ TeV and has a slope in the $\sim 2-15$ TeV energy range 
consistent with what IceCube has observed [$\gamma=3.2\pm0.2$, where $n(E)\propto E^{-\gamma}$ and $n(E)$ is the neutrino particle spectrum; \citealt{Abbasi_2022}]. 

As for the relative normalization between neutrino and X-ray AGN spectra, we resorted to the observed X-ray and neutrino fluxes (or luminosities) of NGC1068, the only non-jetted AGN for which these two crucial parameters are available. We remark that what matters here is the $intrinsic$, i.e. corrected for absorption, AGN X-ray flux (or luminosity), since the total background is obtained by integrating $intrinsic$ AGN luminosity functions. The best estimate of the intrinsic X-ray flux of NGC1068 is arguably the one presented by \cite{Marinucci_2016}, who, by means of NuSTAR observations, obtained spectra of the target up to $\sim60$ keV at different epochs and discovered a transient decrease of the column density along the line of sight, from $\gtrsim10^{25}$ to $7\times 10^{24}$ cm$^{-2}$, which was sufficient to reveal its direct nuclear radiation.
The intrinsic flux density at 1 keV was found to be $f_{1\;keV}=1.44 \times 10^{-9}$ \cgskev. 
We then derived the all-flavour neutrino flux density at 4 TeV, where it is likely best constrained, by considering the muon-type neutrino spectrum measured by IceCube, $EF(E)= 5\times 10^{-11} (E/1{\rm TeV})^{-1.2}$  TeV cm$^{-2}$ s$^{-1}$, and multiplying it by 3 (which assumes vacuum neutrino mixing). This gives a neutrino flux density of $f_{4\;TeV}=1.14\times10^{-20}$ \cgskev , and hence an X-ray to neutrino flux ratio $\nu f_\nu|_{1\; keV}/\nu f_\nu|_{4\; TeV}=31.5$. 
The uncertainty on this value, taking into account the errors 
on the X-ray and neutrino fluxes and those on the photon indices, 
is $\sim 0.5$ dex. 
We used this ratio as the relative normalization between the neutrino and X-ray AGN spectra and computed the expected neutrino background spectrum as:
\begin{equation}
F(E)=\frac{1}{4\pi}\int_{0}^{z_{max}}\frac{1+z}{4\pi {d_L}^2}\;
\frac{dV}{dz}\int_{L_{min}}^{L_{max}}f[E(1+z)]\Phi[L,z]L\,dL\;dz \;,
\end{equation}
where $d_L$ is the luminosity distance, $\frac{dV}{dz}$ is the comoving volume element,  $\Phi[L,z]$ is the total, intrinsic comoving AGN XLF (obtained by summing both obscured and unobscured AGN) as per the CXB model of \citet{Gilli_2007} and $f[E(1+z)]$ is the X-ray normalised neutrino flux density at the energy $E(1+z)$. We integrated the AGN XLF in the luminosity range $L_{0.5-2 keV}=10^{42}-10^{48}$ erg s$^{-1}$ and in the redshift interval $z=0-5$. We note that the contribution of AGN at $z>5$ to the total background flux is negligible and that integrating the AGN XLF down to $L_{0.5-2 keV}=10^{41}$ erg s$^{-1}$ to include the contribution of low-luminosity AGN would increase the total background flux by at most $\sim$16\%. 
For comparison, we also computed the neutrino diffuse background expected from very local AGN, i.e. within the same distance to NGC1068. 


\begin{figure*}
\centering
  \includegraphics[width=17cm]{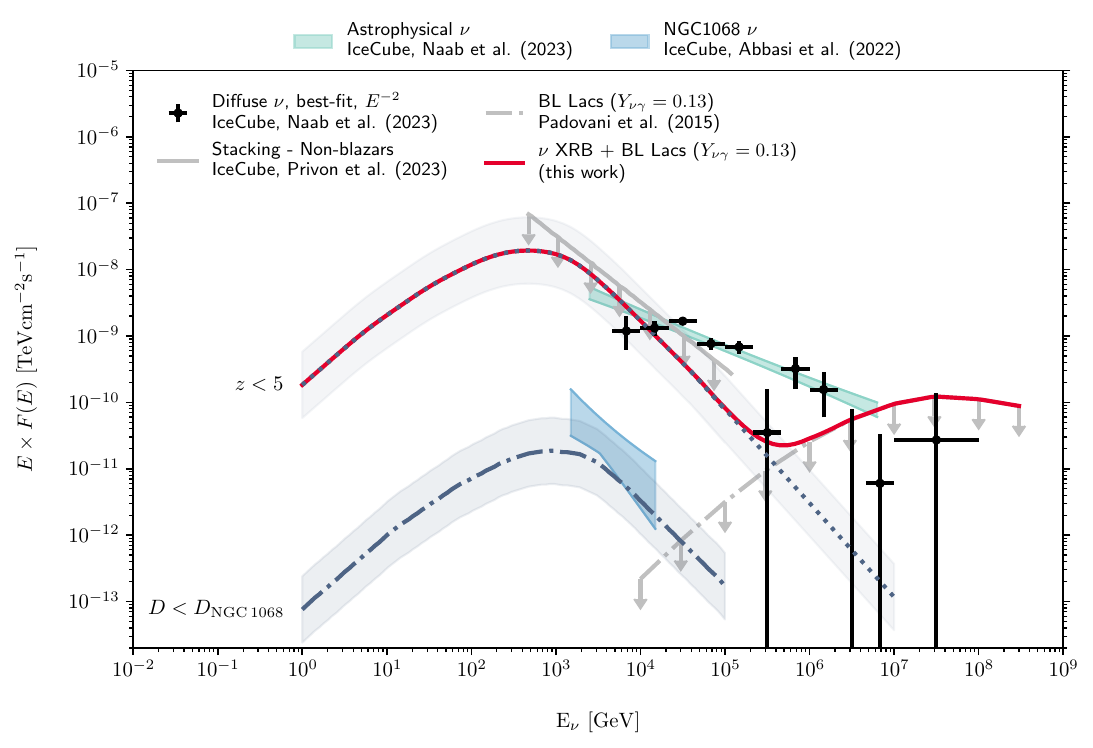}
    \caption{Computed all-flavour neutrino background derived from  an X-ray AGN population synthesis. Dark blue curves show the computed neutrino backgrounds for source populations integrated up to the distance of NGC1068 and redshift $z=5$ (dash-dotted and dotted, respectively). A high-energy extrapolation up to $10^7\,$GeV is added to the integrated spectrum for $z=5$ and combined with the blazar neutrino background model by \cite{Padovani_2015} (dash-dotted grey curve) to highlight the structure of the combined AGN neutrino background flux (``double-humped'' red solid curve). The estimated uncertainty on the integrated neutrino component from X-ray AGN is assumed to be $0.5\,$dex (dark blue band). 
    Also shown are the current best-fit astrophysical diffuse neutrino flux and the segmented neutrino flux fit assuming an $E^{-2}$ energy spectrum in each bin (green area and black points: \citealt{naab2023measurement}), IceCube upper limits from stacking analyses for non-blazar AGN (grey solid line: \citealt{IceCube:2023_privon}), and the point-source neutrino flux of NGC1068 (blue area: \citealt{ngc1068}).}
    \label{fig:diffuse_w_GP}
\end{figure*}

\section{Main results}

Fig. \ref{fig:diffuse_w_GP} shows the resulting all-flavour neutrino background integrated up 
to redshift $z=5$ (dotted blue curve). This can be compared with the current best-fit for the 
IceCube astrophysical diffuse neutrino flux as obtained by \cite{naab2023measurement},
derived using a segmented neutrino flux fit with individual energy bins assuming an $E^{-2}$ 
energy spectrum in each bin (black points) and a single power-law fit $\propto E^{-2.52\pm0.04}$ 
(green area). It can be seen that our integrated non-jetted AGN contribution is consistent
with the IceCube diffuse component down to $\approx 10$ TeV. In other words, the assumption that 
all non-jetted AGN behave like NGC 1068 in terms of their neutrino properties is not inconsistent 
with present IceCube data above this energy. 

Most likely, different astrophysical populations contribute to the IceCube diffuse and hence
this comparison might be not be very constraining. In Fig. \ref{fig:diffuse_w_GP}, therefore, we
also show upper limits derived by IceCube from a stacking analysis carried out on 
non-blazar (and therefore mostly non-jetted) AGN (grey solid line: \citealt{IceCube:2023_privon}). 
The analysis tested the hypothesis of neutrino emission from hard X-ray AGN in the BASS catalogue \citep{Ricci_2017}, assuming that the neutrino flux correlates with the de-absorbed X-ray one. After finding no significant results, \cite{IceCube:2023_privon} placed 90\% C.L. upper limits on the neutrino emission from all non-jetted hard X-ray AGN assuming an energy spectrum $\propto E_{\nu}^{-3}$. These limits were corrected for catalogue incompleteness by accounting for the missing neutrino signal from all non-detected non-jetted X-ray AGN in the Universe using the XLF given in \cite{Ueda_2014}.
Our results turn out to be not inconsistent even with the estimated maximum emission from
non-jetted AGN, although barely so. 

Fig. \ref{fig:diffuse_w_GP} shows also the diffuse neutrino background expected from all AGN 
within 10.1 Mpc, that is the most reliable distance to NGC1068 (Section \ref{sec:intro}: dash-dotted blue curve). 
At first glance this is somewhat smaller than 
the neutrino flux from NGC1068 alone. We note however that the 0.5 dex uncertainty in the X-ray 
to neutrino flux ratio assumed for NGC1068 (Section \ref{sec:predict}) alleviates significantly 
this tension. Moreover, the model predictions may suffer from significant limitations when tiny 
redshift (distance) intervals are considered, primarily because of statistical fluctuations. 
The shape and evolution of the AGN XLFs are in fact derived from samples of thousands of objects 
distributed up to large cosmological distances ($z\lesssim5$). By contrast, only three AGN with 
intrinsic $L_{0.5-2 keV}>10^{42}$ erg s$^{-1}$ (the luminosity limit used in our integrations), 
including NGC 1068, fall within 10.1 Mpc, as derived from the BASS DR2 catalogue 
\citep{Koss_2022}. 
The number of sources predicted by the assumed XLFs within this small volume is then bound to
be somewhat inaccurate, whereas it becomes more precise, up to a few percent level, when 
integrating over the full redshift range (dotted blue curve in Fig.\ref{fig:diffuse_w_GP}).

Our results are crucially model-independent.
 This means that {\it any} model capable of reproducing the IceCube data for NGC 1068 would yield very similar curves in the energy range covered by current neutrino data. Consequently, our conclusions hold regardless of the specific mechanism driving neutrino emission in this source.

To try to get the bigger picture we added a high-energy extrapolation up to $10^7\,$GeV to 
the whole AGN integrated spectrum and then combined it with the blazar 
neutrino background model by \cite{Padovani_2015}\footnote{As explained in 
\cite{Padovani_2022}, the blazar curve is 
scaled down by a factor $\sim 6.2$ as compared to the original one not to violate the upper 
limits of \cite{Aartsen_2016}.} (dash-dotted grey curve). Our overall results are shown by the 
``double-humped'' red solid curve and present a possible scenario where non-jetted AGN
contribute mostly to the low-energy ($\lesssim$ 1 PeV) IceCube diffuse whereas blazars dominate
the high-energy part. This would be in tantalising agreement with the dip in the data at 
$\sim 300\,$TeV, which might then be related to the fall of the non-jetted AGN contribution
and the rise of the blazar one. 

Can we improve on our predictions? Preliminary results indicate that the IceCube data associated 
with the selection of Seyfert galaxies in the Northern Sky, in particular NGC 4151 and CGCG 
420-015, are inconsistent with the neutrino background at the $2.7\,\sigma$ level of significance 
\citep{IceCube:2023_qinrui}. Moreover, the IceCube search for high-energy neutrino emission from hard
X-ray AGN has reported NGC 4151 at a
significance level of $2.9\,\sigma$ level \citep{IceCube:2023_privon}.
Using the IceCube fluxes and spectral slopes for these two sources \citep{TAUP2023} 
and typical X-ray data (powers and spectra) from \cite{Wang_2010} and \cite{Tanimoto_2022} respectively, we derive
X-ray to neutrino flux ratios $\nu f_\nu|_{1\; keV}/\nu f_\nu|_{4\; TeV}
\sim 5.4$ and $\sim 0.5$, i.e. $\sim 6$ and $\sim 60$ times lower than
for NGC 1068. At face value this would then mean that NGC 1068 has a lower than average
neutrino to X-ray ratio, implying that the dotted blue curve in Fig. 
\ref{fig:diffuse_w_GP} should be shifted upwards by $\sim 1 - 2$ orders of magnitude. As this 
would violate by the same amount the AGN IceCube upper limits from 
\citet{IceCube:2023_privon}, it would then follow that only $\sim 1 - 10\%$ of non-jetted AGN are neutrino 
emitters. Clearly, this is an issue for which more and better IceCube data are of paramount 
importance, both on the source and population side. 

In summary, a population synthesis model, which accounts for the X-ray background from non-jetted
AGN, predicts a neutrino background not violating current IceCube data and upper limits, when using 
NGC 1068 to convert from one band to the other. Preliminary results from two more Seyferts appear
instead to imply an over-prediction. Further IceCube observations are obviously needed to sort out this
issue.

\begin{acknowledgements}
We thank Stefano Gabici and Bj\"orn Eichmann for useful comments. 
This work is supported by the Deutsche Forschungsgemeinschaft (DFG, German Research Foundation) through grant SFB 1258 ``Neutrinos and Dark Matter in Astro- and Particle Physics''.
\end{acknowledgements}

%
   \bibliographystyle{aa} 
   \bibliography{AGN_diffuse_revised_arxiv} 

\begin{thebibliography}{38}
\expandafter\ifx\csname natexlab\endcsname\relax\def\natexlab#1{#1}\fi

\bibitem[{{Aartsen} {et~al.}(2016){Aartsen}, {Abraham}, {Ackermann}, {Adams},
  {Aguilar}, {Ahlers}, {Ahrens}, {Altmann}, {Andeen}, {Anderson}, {Ansseau},
  {Anton}, {Archinger}, {Arg{\"u}elles}, {Auffenberg}, {Axani}, {Bai},
  {Barwick}, {Baum}, {Bay}, {Beatty}, {Becker Tjus}, {Becker}, {BenZvi},
  {Berghaus}, {Berley}, {Bernardini}, {Bernhard}, {Besson}, {Binder}, {Bindig},
  {Bissok}, {Blaufuss}, {Blot}, {Bohm}, {B{\"o}rner}, {Bos}, {Bose},
  {B{\"o}ser}, {Botner}, {Braun}, {Brayeur}, {Bretz}, {Burgman}, {Carver},
  {Casier}, {Cheung}, {Chirkin}, {Christov}, {Clark}, {Classen}, {Coenders},
  {Collin}, {Conrad}, {Cowen}, {Cross}, {Day}, {de Andr{\'e}}, {De Clercq},
  {del Pino Rosendo}, {Dembinski}, {De Ridder}, {Desiati}, {de Vries}, {de
  Wasseige}, {de With}, {DeYoung}, {D{\'\i}az-V{\'e}lez}, {di Lorenzo},
  {Dujmovic}, {Dumm}, {Dunkman}, {Eberhardt}, {Ehrhardt}, {Eichmann}, {Eller},
  {Euler}, {Evenson}, {Fahey}, {Fazely}, {Feintzeig}, {Felde}, {Filimonov},
  {Finley}, {Flis}, {F{\"o}sig}, {Franckowiak}, {Friedman}, {Fuchs}, {Gaisser},
  {Gallagher}, {Gerhardt}, {Ghorbani}, {Giang}, {Gladstone}, {Glagla},
  {Gl{\"u}senkamp}, {Goldschmidt}, {Golup}, {Gonzalez}, {Grant}, {Griffith},
  {Haack}, {Haj Ismail}, {Hallgren}, {Halzen}, {Hansen}, {Hansmann},
  {Hansmann}, {Hanson}, {Hebecker}, {Heereman}, {Helbing}, {Hellauer},
  {Hickford}, {Hignight}, {Hill}, {Hoffman}, {Hoffmann}, {Holzapfel},
  {Hoshina}, {Huang}, {Huber}, {Hultqvist}, {In}, {Ishihara}, {Jacobi},
  {Japaridze}, {Jeong}, {Jero}, {Jones}, {Jurkovic}, {Kappes}, {Karg}, {Karle},
  {Katz}, {Kauer}, {Keivani}, {Kelley}, {Kemp}, {Kheirandish}, {Kim},
  {Kintscher}, {Kiryluk}, {Kittler}, {Klein}, {Kohnen}, {Koirala}, {Kolanoski},
  {Konietz}, {K{\"o}pke}, {Kopper}, {Kopper}, {Koskinen}, {Kowalski}, {Krings},
  {Kroll}, {Kr{\"u}ckl}, {Kr{\"u}ger}, {Kunnen}, {Kunwar}, {Kurahashi},
  {Kuwabara}, {Labare}, {Lanfranchi}, {Larson}, {Lauber}, {Lennarz},
  {Lesiak-Bzdak}, {Leuermann}, {Leuner}, {Lu}, {L{\"u}nemann}, {Madsen},
  {Maggi}, {Mahn}, {Mancina}, {Mandelartz}, {Maruyama}, {Mase}, {Maunu},
  {McNally}, {Meagher}, {Medici}, {Meier}, {Meli}, {Menne}, {Merino}, {Meures},
  {Miarecki}, {Mohrmann}, {Montaruli}, {Moulai}, {Nahnhauer}, {Naumann},
  {Neer}, {Niederhausen}, {Nowicki}, {Nygren}, {Obertacke Pollmann}, {Olivas},
  {O'Murchadha}, {Palczewski}, {Pandya}, {Pankova}, {Penek}, {Pepper},
  {P{\'e}rez de los Heros}, {Pieloth}, {Pinat}, {Price}, {Przybylski},
  {Quinnan}, {Raab}, {R{\"a}del}, {Rameez}, {Rawlins}, {Reimann}, {Relethford},
  {Relich}, {Resconi}, {Rhode}, {Richman}, {Riedel}, {Robertson}, {Rongen},
  {Rott}, {Ruhe}, {Ryckbosch}, {Rysewyk}, {Sabbatini}, {Sanchez Herrera},
  {Sandrock}, {Sandroos}, {Sarkar}, {Satalecka}, {Schimp}, {Schlunder},
  {Schmidt}, {Schoenen}, {Sch{\"o}neberg}, {Schumacher}, {Seckel}, {Seunarine},
  {Soldin}, {Song}, {Spiczak}, {Spiering}, {Stahlberg}, {Stanev}, {Stasik},
  {Steuer}, {Stezelberger}, {Stokstad}, {St{\"o}{\ss}l}, {Str{\"o}m},
  {Strotjohann}, {Sullivan}, {Sutherland}, {Taavola}, {Taboada}, {Tatar},
  {Tenholt}, {Ter-Antonyan}, {Terliuk}, {Te{\v{s}}i{\'c}}, {Tilav}, {Toale},
  {Tobin}, {Toscano}, {Tosi}, {Tselengidou}, {Turcati}, {Unger}, {Usner},
  {Vandenbroucke}, {van Eijndhoven}, {Vanheule}, {van Rossem}, {van Santen},
  {Veenkamp}, {Vehring}, {Voge}, {Vraeghe}, {Walck}, {Wallace}, {Wallraff},
  {Wandkowsky}, {Weaver}, {Weiss}, {Wendt}, {Westerhoff}, {Whelan}, {Wickmann},
  {Wiebe}, {Wiebusch}, {Wille}, {Williams}, {Wills}, {Wolf}, {Wood}, {Woolsey},
  {Woschnagg}, {Xu}, {Xu}, {Xu}, {Yanez}, {Yodh}, {Yoshida}, {Zoll}, \&
  {IceCube Collaboration}}]{Aartsen_2016}
{Aartsen}, M.~G., {Abraham}, K., {Ackermann}, M., {et~al.} 2016, \prl, 117,
  241101

\bibitem[{Abbasi {et~al.}(2022)Abbasi, Ackermann, Adams, Aguilar, Ahlers,
  Ahrens, Alameddine, Alispach, Alves, Amin, Andeen, Anderson, Anton,
  Argüelles, Ashida, Axani, Bai, Balagopal~V., Barbano, Barwick, Bastian,
  Basu, Baur, Bay, Beatty, Becker, Becker~Tjus, Bellenghi, BenZvi, Berley,
  Bernardini, Besson, Binder, Bindig, Blaufuss, Blot, Boddenberg, Bontempo,
  Borowka, Böser, Botner, Böttcher, Bourbeau, Bradascio, Braun, Brinson,
  Bron, Brostean-Kaiser, Browne, Burgman, Burley, Busse, Campana,
  Carnie-Bronca, Chen, Chen, Chirkin, Choi, Clark, Clark, Classen, Coleman,
  Collin, Conrad, Coppin, Correa, Cowen, Cross, Dappen, Dave, De~Clercq,
  DeLaunay, Delgado~López, Dembinski, Deoskar, Desai, Desiati, de~Vries,
  de~Wasseige, de~With, DeYoung, Diaz, Díaz-Vélez, Dittmer, Dujmovic,
  Dunkman, DuVernois, Dvorak, Ehrhardt, Eller, Engel, Erpenbeck, Evans,
  Evenson, Fan, Fazely, Fedynitch, Feigl, Fiedlschuster, Fienberg, Filimonov,
  Finley, Fischer, Fox, Franckowiak, Friedman, Fritz, Fürst, Gaisser,
  Gallagher, Ganster, Garcia, Garrappa, Gerhardt, Ghadimi, Glaser, Glauch,
  Glüsenkamp, Goldschmidt, Gonzalez, Goswami, Grant, Grégoire, Griswold,
  Günther, Gutjahr, Haack, Hallgren, Halliday, Halve, Halzen, Ha~Minh, Hanson,
  Hardin, Harnisch, Haungs, Hebecker, Helbing, Henningsen, Hettinger, Hickford,
  Hignight, Hill, Hill, Hoffman, Hoffmann, Hokanson-Fasig, Hoshina, Huang,
  Huber, Huber, Hultqvist, Hünnefeld, Hussain, Hymon, In, Iovine, Ishihara,
  Jansson, Japaridze, Jeong, Jin, Jones, Kang, Kang, Kang, Kappes, Kappesser,
  Kardum, Karg, Karl, Karle, Katz, Kauer, Kellermann, Kelley, Kheirandish, Kin,
  Kintscher, Kiryluk, Klein, Koirala, Kolanoski, Kontrimas, Köpke, Kopper,
  Kopper, Koskinen, Koundal, Kovacevich, Kowalski, Kozynets, Kun, Kurahashi,
  Lad, Lagunas~Gualda, Lanfranchi, Larson, Lauber, Lazar, Lee, Leonard,
  Leszczyńska, Li, Lincetto, Liu, Liubarska, Lohfink, Lozano~Mariscal, Lu,
  Lucarelli, Ludwig, Luszczak, Lyu, Ma, Madsen, Mahn, Makino, Mancina, Mariş,
  Martinez-Soler, Maruyama, Mase, McElroy, McNally, Mead, Meagher, Mechbal,
  Medina, Meier, Meighen-Berger, Micallef, Mockler, Montaruli, Moore, Morse,
  Moulai, Naab, Nagai, Nahnhauer, Naumann, Necker, Nguyen, Niederhausen, Nisa,
  Nowicki, Nygren, Obertacke~Pollmann, Oehler, Oeyen, Olivas, O’Sullivan,
  Pandya, Pankova, Park, Parker, Paudel, Paul, Pérez de~los Heros, Peters,
  Peterson, Philippen, Pieper, Pittermann, Pizzuto, Plum, Popovych, Porcelli,
  Prado~Rodriguez, Price, Pries, Przybylski, Raab, Rack-Helleis, Raissi,
  Rameez, Rawlins, Rea, Rehman, Reichherzer, Reimann, Renzi, Resconi, Reusch,
  Rhode, Richman, Riedel, Roberts, Robertson, Roellinghoff, Rongen, Rott, Ruhe,
  Ryckbosch, Rysewyk~Cantu, Safa, Saffer, Sanchez~Herrera, Sandrock, Sandroos,
  Santander, Sarkar, Sarkar, Satalecka, Schaufel, Schieler, Schindler, Schmidt,
  Schneider, Schneider, Schröder, Schumacher, Schwefer, Sclafani, Seckel,
  Seunarine, Sharma, Shefali, Silva, Skrzypek, Smithers, Snihur, Soedingrekso,
  Soldin, Spannfellner, Spiczak, Spiering, Stachurska, Stamatikos, Stanev,
  Stein, Stettner, Steuer, Stezelberger, Stokstad, Stürwald, Stuttard,
  Sullivan, Taboada, Ter-Antonyan, Tilav, Tischbein, Tollefson, Tönnis,
  Toscano, Tosi, Trettin, Tselengidou, Tung, Turcati, Turcotte, Turley,
  Twagirayezu, Ty, Unland~Elorrieta, Valtonen-Mattila, Vandenbroucke, van
  Eijndhoven, Vannerom, van Santen, Verpoest, Walck, Watson, Weaver, Weigel,
  Weindl, Weiss, Weldert, Wendt, Werthebach, Weyrauch, Whitehorn, Wiebusch,
  Williams, Wolf, Woschnagg, Wrede, Wulff, Xu, Yanez, Yoshida, Yu, Yuan, Zhang,
  \& Zhelnin}]{ngc1068}
Abbasi, R., Ackermann, M., Adams, J., {et~al.} 2022, Science, 378, 538

\bibitem[{Abbasi {et~al.}(2023)}]{IceCube:2023_qinrui}
Abbasi, R. {et~al.} 2023, PoS, ICRC2023, 1052

\bibitem[{{Akylas} {et~al.}(2012){Akylas}, {Georgakakis}, {Georgantopoulos},
  {Brightman}, \& {Nandra}}]{Akylas_2012}
{Akylas}, A., {Georgakakis}, A., {Georgantopoulos}, I., {Brightman}, M., \&
  {Nandra}, K. 2012, \aap, 546, A98

\bibitem[{{Ananna} {et~al.}(2019){Ananna}, {Treister}, {Urry}, {Ricci},
  {Kirkpatrick}, {LaMassa}, {Buchner}, {Civano}, {Tremmel}, \&
  {Marchesi}}]{Ananna_2019}
{Ananna}, T.~T., {Treister}, E., {Urry}, C.~M., {et~al.} 2019, \apj, 871, 240

\bibitem[{{Berezinsky}(1977)}]{berezinsky1977}
{Berezinsky}, V.~S. 1977, {Proceedings of the International Conference Neutrino
  '77}, 177

\bibitem[{{Comastri} {et~al.}(1995){Comastri}, {Setti}, {Zamorani}, \&
  {Hasinger}}]{Comastri_1995}
{Comastri}, A., {Setti}, G., {Zamorani}, G., \& {Hasinger}, G. 1995, \aap, 296,
  1

\bibitem[{{Eichler}(1979)}]{eichler1979}
{Eichler}, D. 1979, \apj, 232, 106

\bibitem[{{Gilli} {et~al.}(2007){Gilli}, {Comastri}, \&
  {Hasinger}}]{Gilli_2007}
{Gilli}, R., {Comastri}, A., \& {Hasinger}, G. 2007, \aap, 463, 79

\bibitem[{{Giommi} \& {Padovani}(2021)}]{Giommmi_2021}
{Giommi}, P. \& {Padovani}, P. 2021, Universe, 7, 492

\bibitem[{{Hasinger} {et~al.}(2005){Hasinger}, {Miyaji}, \&
  {Schmidt}}]{HMS_2005}
{Hasinger}, G., {Miyaji}, T., \& {Schmidt}, M. 2005, \aap, 441, 417

\bibitem[{{IceCube Collaboration} {et~al.}(2022){IceCube Collaboration},
  {Abbasi}, {Ackermann}, {Adams}, {Aguilar}, {Ahlers}, {Ahrens}, {Alameddine},
  {Alispach}, {Alves}, {Amin}, {Andeen}, {Anderson}, {Anton}, {Arg{\"u}elles},
  {Ashida}, {Axani}, {Bai}, {Balagopal}, {Barbano}, {Barwick}, {Bastian},
  {Basu}, {Baur}, {Bay}, {Beatty}, {Becker}, {Becker Tjus}, {Bellenghi},
  {Benzvi}, {Berley}, {Bernardini}, {Besson}, {Binder}, {Bindig}, {Blaufuss},
  {Blot}, {Boddenberg}, {Bontempo}, {Borowka}, {B{\"o}ser}, {Botner},
  {B{\"o}ttcher}, {Bourbeau}, {Bradascio}, {Braun}, {Brinson}, {Bron},
  {Brostean-Kaiser}, {Browne}, {Burgman}, {Burley}, {Busse}, {Campana},
  {Carnie-Bronca}, {Chen}, {Chen}, {Chirkin}, {Choi}, {Clark}, {Clark},
  {Classen}, {Coleman}, {Collin}, {Conrad}, {Coppin}, {Correa}, {Cowen},
  {Cross}, {Dappen}, {Dave}, {de Clercq}, {Delaunay}, {Delgado L{\'o}pez},
  {Dembinski}, {Deoskar}, {Desai}, {Desiati}, {de Vries}, {de Wasseige}, {de
  With}, {Deyoung}, {Diaz}, {D{\'\i}az-V{\'e}lez}, {Dittmer}, {Dujmovic},
  {Dunkman}, {Duvernois}, {Dvorak}, {Ehrhardt}, {Eller}, {Engel}, {Erpenbeck},
  {Evans}, {Evenson}, {Fan}, {Fazely}, {Fedynitch}, {Feigl}, {Fiedlschuster},
  {Fienberg}, {Filimonov}, {Finley}, {Fischer}, {Fox}, {Franckowiak},
  {Friedman}, {Fritz}, {F{\"u}rst}, {Gaisser}, {Gallagher}, {Ganster},
  {Garcia}, {Garrappa}, {Gerhardt}, {Ghadimi}, {Glaser}, {Glauch},
  {Gl{\"u}senkamp}, {Goldschmidt}, {Gonzalez}, {Goswami}, {Grant},
  {Gr{\'e}goire}, {Griswold}, {G{\"u}nther}, {Gutjahr}, {Haack}, {Hallgren},
  {Halliday}, {Halve}, {Halzen}, {Hanson}, {Hardin}, {Harnisch}, {Haungs},
  {Hebecker}, {Helbing}, {Henningsen}, {Hettinger}, {Hickford}, {Hignight},
  {Hill}, {Hill}, {Hoffman}, {Hoffmann}, {Hokanson-Fasig}, {Hoshina}, {Huang},
  {Huber}, {Huber}, {Hultqvist}, {H{\"u}nnefeld}, {Hussain}, {Hymon}, {in},
  {Iovine}, {Ishihara}, {Jansson}, {Japaridze}, {Jeong}, {Jin}, {Jones},
  {Kang}, {Kang}, {Kang}, {Kappes}, {Kappesser}, {Kardum}, {Karg}, {Karl},
  {Karle}, {Katz}, {Kauer}, {Kellermann}, {Kelley}, {Kheirandish}, {Kin},
  {Kintscher}, {Kiryluk}, {Klein}, {Koirala}, {Kolanoski}, {Kontrimas},
  {K{\"o}pke}, {Kopper}, {Kopper}, {Koskinen}, {Koundal}, {Kovacevich},
  {Kowalski}, {Kozynets}, {Kun}, {Kurahashi}, {Lad}, {Lagunas Gualda},
  {Lanfranchi}, {Larson}, {Lauber}, {Lazar}, {Lee}, {Leonard},
  {Leszczy{\'n}ska}, {Li}, {Lincetto}, {Liu}, {Liubarska}, {Lohfink}, {Lozano
  Mariscal}, {Lu}, {Lucarelli}, {Ludwig}, {Luszczak}, {Lyu}, {Ma}, {Madsen},
  {Mahn}, {Makino}, {Mancina}, {Mari{\c{s}}}, {Martinez-Soler}, {Maruyama},
  {Mase}, {McElroy}, {McNally}, {Mead}, {Meagher}, {Mechbal}, {Medina},
  {Meier}, {Meighen-Berger}, {Micallef}, {Mockler}, {Montaruli}, {Moore},
  {Morse}, {Moulai}, {Naab}, {Nagai}, {Nahnhauer}, {Naumann}, {Necker},
  {Nguyen}, {Niederhausen}, {Nisa}, {Nowicki}, {Nygren}, {Obertack},
  {Pollmann}, {Oehler}, {Oeyen}, {Olivas}, {O'Sullivan}, {Pandya}, {Pankova},
  {Park}, {Parker}, {Paudel}, {Paul}, {P{\'e}rez de Los Heros}, {Peters},
  {Peterson}, {Philippen}, {Pieper}, {Pittermann}, {Pizzuto}, {Plum},
  {Popovych}, {Porcelli}, {Prado Rodriguez}, {Price}, {Pries}, {Przybylski},
  {Rack-Helleis}, {Raissi}, {Rameez}, {Rawlins}, {Rea}, {Rehman},
  {Reichherzer}, {Reimann}, {Renzi}, {Resconi}, {Reusch}, {Rhode}, {Richman},
  {Riedel}, {Roberts}, {Robertson}, {Roellinghoff}, {Rongen}, {Rott}, {Ruhe},
  {Ryckbosch}, {Rysewyk Cantu}, {Safa}, {Saffer}, {Sanchez Herrera},
  {Sandrock}, {Sandroos}, {Santander}, {Sarkar}, {Sarkar}, {Satalecka},
  {Schaufel}, {Schieler}, {Schindler}, {Schmidt}, {Schneider}, {Schneider},
  {Schr{\"o}der}, {Schumacher}, {Schwefer}, {Sclafani}, {Seckel}, {Seunarine},
  {Sharma}, {Shefali}, {Silva}, {Skrzypek}, {Smithers}, {Snihur},
  {Soedingrekso}, {Soldin}, {Spannfellner}, {Spiczak}, {Spiering},
  {Stachurska}, {Stamatikos}, {Stanev}, {Stein}, {Stettner}, {Steuer},
  {Stezelberger}, {Stokstad}, {St{\"u}rwald}, {Stuttard}, {Sullivan},
  {Taboada}, {Ter-Antonyan}, {Tilav}, {Tischbein}, {Tollefson}, {T{\"o}nnis},
  {Toscano}, {Tosi}, {Trettin}, {Tselengidou}, {Tung}, {Turcati}, {Turcotte},
  {Turley}, {Twagirayezu}, {Ty}, {Unland Elorrieta}, {Valtonen-Mattila},
  {Vandenbroucke}, {van Eijndhoven}, {Vannerom}, {van Santen}, {Verpoest},
  {Walck}, {Watson}, {Weaver}, {Weigel}, {Weindl}, {Weiss}, {Weldert}, {Wendt},
  {Werthebach}, {Weyrauch}, {Whitehorn}, {Wiebusch}, {Williams}, {Wolf},
  {Woschnagg}, {Wrede}, {Wulff}, {Xu}, {Yanez}, {Yoshida}, {Yu}, {Yuan},
  {Zhangan}, \& {Zhelnin}}]{Abbasi_2022}
{IceCube Collaboration}, {Abbasi}, R., {Ackermann}, M., {et~al.} 2022, Science,
  378, 538

\bibitem[{Kheirandish {et~al.}(2021)Kheirandish, Murase, \&
  Kimura}]{Kheirandish_2021}
Kheirandish, A., Murase, K., \& Kimura, S.~S. 2021, Astrophys. J., 922, 45

\bibitem[{{Koss} {et~al.}(2022){Koss}, {Ricci}, {Trakhtenbrot}, {Oh}, {den
  Brok}, {Mej{\'\i}a-Restrepo}, {Stern}, {Privon}, {Treister}, {Powell},
  {Mushotzky}, {Bauer}, {Ananna}, {Balokovi{\'c}}, {B{\"a}r}, {Becker},
  {Bessiere}, {Burtscher}, {Caglar}, {Congiu}, {Evans}, {Harrison}, {Heida},
  {Ichikawa}, {Kamraj}, {Lamperti}, {Pacucci}, {Ricci}, {Riffel}, {Rojas},
  {Schawinski}, {Temple}, {Urry}, {Veilleux}, \& {Williams}}]{Koss_2022}
{Koss}, M.~J., {Ricci}, C., {Trakhtenbrot}, B., {et~al.} 2022, \apjs, 261, 2

\bibitem[{{La Franca} {et~al.}(2005){La Franca}, {Fiore}, {Comastri}, {Perola},
  {Sacchi}, {Brusa}, {Cocchia}, {Feruglio}, {Matt}, {Vignali}, {Carangelo},
  {Ciliegi}, {Lamastra}, {Maiolino}, {Mignoli}, {Molendi}, \&
  {Puccetti}}]{Lafranca_2005}
{La Franca}, F., {Fiore}, F., {Comastri}, A., {et~al.} 2005, \apj, 635, 864

\bibitem[{{Luo} {et~al.}(2017){Luo}, {Brandt}, {Xue}, {Lehmer}, {Alexander},
  {Bauer}, {Vito}, {Yang}, {Basu-Zych}, {Comastri}, {Gilli}, {Gu},
  {Hornschemeier}, {Koekemoer}, {Liu}, {Mainieri}, {Paolillo}, {Ranalli},
  {Rosati}, {Schneider}, {Shemmer}, {Smail}, {Sun}, {Tozzi}, {Vignali}, \&
  {Wang}}]{Luo_2017}
{Luo}, B., {Brandt}, W.~N., {Xue}, Y.~Q., {et~al.} 2017, \apjs, 228, 2

\bibitem[{{Marchesi} {et~al.}(2020){Marchesi}, {Gilli}, {Lanzuisi}, {Dauser},
  {Ettori}, {Vito}, {Cappelluti}, {Comastri}, {Mushotzky}, {Ptak}, \&
  {Norman}}]{Marchesi_2020}
{Marchesi}, S., {Gilli}, R., {Lanzuisi}, G., {et~al.} 2020, \aap, 642, A184

\bibitem[{{Marinucci} {et~al.}(2016){Marinucci}, {Bianchi}, {Matt},
  {Alexander}, {Balokovi{\'c}}, {Bauer}, {Brandt}, {Gandhi}, {Guainazzi},
  {Harrison}, {Iwasawa}, {Koss}, {Madsen}, {Nicastro}, {Puccetti}, {Ricci},
  {Stern}, \& {Walton}}]{Marinucci_2016}
{Marinucci}, A., {Bianchi}, S., {Matt}, G., {et~al.} 2016, \mnras, 456, L94

\bibitem[{{Murase}(2022)}]{Murase_2022}
{Murase}, K. 2022, \apjl, 941, L17

\bibitem[{Naab {et~al.}(2023)Naab, Ganster, \& Zhang}]{naab2023measurement}
Naab, R., Ganster, E., \& Zhang, Z. 2023, PoS, ICRC2023, 1064

\bibitem[{{Nanni} {et~al.}(2020){Nanni}, {Gilli}, {Vignali}, {Mignoli}, {Peca},
  {Marchesi}, {Annunziatella}, {Brusa}, {Calura}, {Cappelluti}, {Chiaberge},
  {Comastri}, {Iwasawa}, {Lanzuisi}, {Liuzzo}, {Marchesini}, {Prandoni},
  {Tozzi}, {Vito}, {Zamorani}, \& {Norman}}]{Nanni_2020}
{Nanni}, R., {Gilli}, R., {Vignali}, C., {et~al.} 2020, \aap, 637, A52

\bibitem[{{Padovani}(2017)}]{Padovani_2017NatAs}
{Padovani}, P. 2017, Nature Astronomy, 1, 0194

\bibitem[{{Padovani} {et~al.}(2017){Padovani}, {Alexander}, {Assef}, {De
  Marco}, {Giommi}, {Hickox}, {Richards}, {Smol{\v{c}}i{\'c}},
  {Hatziminaoglou}, {Mainieri}, \& {Salvato}}]{Padovani_2017}
{Padovani}, P., {Alexander}, D.~M., {Assef}, R.~J., {et~al.} 2017, \aapr, 25, 2

\bibitem[{{Padovani} {et~al.}(2022){Padovani}, {Giommi}, {Falomo}, {Oikonomou},
  {Petropoulou}, {Glauch}, {Resconi}, {Treves}, \& {Paiano}}]{Padovani_2022}
{Padovani}, P., {Giommi}, P., {Falomo}, R., {et~al.} 2022, \mnras, 510, 2671

\bibitem[{Padovani {et~al.}(2015)Padovani, Petropoulou, Giommi, \&
  Resconi}]{Padovani_2015}
Padovani, P., Petropoulou, M., Giommi, P., \& Resconi, E. 2015, MNRAS, 452,
  1877

\bibitem[{{Padovani} {et~al.}(2024){Padovani}, {Resconi}, {Ajello},
  {Bellenghi}, {Bianchi}, {Blasi}, {Huang}, {Gabici}, {G\`amez Rosas},
  {Niederhausen}, {Peretti}, {Eichmann}, {Guetta}, {Lamastra}, \&
  {Shimizu}}]{Padovani_2024}
{Padovani}, P., {Resconi}, E., {Ajello}, M., {et~al.} 2024, Nature Astronomy,
  submitted

\bibitem[{{Privon} {et~al.}(2023)}]{IceCube:2023_privon}
{Privon}, G.~C. {et~al.} 2023, PoS, ICRC2023, 1032

\bibitem[{{{Qinrui}, L. and {IceCube Collaboration}}(2023)}]{TAUP2023}
{{Qinrui}, L. and {IceCube Collaboration}}. 2023, in XVIII International
  Conference on Topics in Astoparticle and Underground Physics (TAUP2023)

\bibitem[{{Ricci} {et~al.}(2017){Ricci}, {Trakhtenbrot}, {Koss}, {Ueda}, {Del
  Vecchio}, {Treister}, {Schawinski}, {Paltani}, {Oh}, {Lamperti}, {Berney},
  {Gandhi}, {Ichikawa}, {Bauer}, {Ho}, {Asmus}, {Beckmann}, {Soldi},
  {Balokovi{\'c}}, {Gehrels}, \& {Markwardt}}]{Ricci_2017}
{Ricci}, C., {Trakhtenbrot}, B., {Koss}, M.~J., {et~al.} 2017, \apjs, 233, 17

\bibitem[{{Risaliti} {et~al.}(1999){Risaliti}, {Maiolino}, \&
  {Salvati}}]{Risaliti_1999}
{Risaliti}, G., {Maiolino}, R., \& {Salvati}, M. 1999, \apj, 522, 157

\bibitem[{{Setti} \& {Woltjer}(1989)}]{Setti_1989}
{Setti}, G. \& {Woltjer}, L. 1989, \aap, 224, L21

\bibitem[{{Silberberg} \& {Shapiro}(1979)}]{silberberg1979}
{Silberberg}, R. \& {Shapiro}, M.~M. 1979, in International Cosmic Ray
  Conference, Vol.~10, International Cosmic Ray Conference, 357

\bibitem[{{Tanimoto} {et~al.}(2022){Tanimoto}, {Ueda}, {Odaka}, {Yamada}, \&
  {Ricci}}]{Tanimoto_2022}
{Tanimoto}, A., {Ueda}, Y., {Odaka}, H., {Yamada}, S., \& {Ricci}, C. 2022,
  \apjs, 260, 30

\bibitem[{{Tozzi} {et~al.}(2006){Tozzi}, {Gilli}, {Mainieri}, {Norman},
  {Risaliti}, {Rosati}, {Bergeron}, {Borgani}, {Giacconi}, {Hasinger},
  {Nonino}, {Streblyanska}, {Szokoly}, {Wang}, \& {Zheng}}]{Tozzi_2006}
{Tozzi}, P., {Gilli}, R., {Mainieri}, V., {et~al.} 2006, \aap, 451, 457

\bibitem[{{Treister} \& {Urry}(2006)}]{Treister_2006}
{Treister}, E. \& {Urry}, C.~M. 2006, \apjl, 652, L79

\bibitem[{{Ueda} {et~al.}(2014){Ueda}, {Akiyama}, {Hasinger}, {Miyaji}, \&
  {Watson}}]{Ueda_2014}
{Ueda}, Y., {Akiyama}, M., {Hasinger}, G., {Miyaji}, T., \& {Watson}, M.~G.
  2014, \apj, 786, 104

\bibitem[{{Ueda} {et~al.}(2003){Ueda}, {Akiyama}, {Ohta}, \&
  {Miyaji}}]{Ueda_2003}
{Ueda}, Y., {Akiyama}, M., {Ohta}, K., \& {Miyaji}, T. 2003, \apj, 598, 886

\bibitem[{{Wang} {et~al.}(2010){Wang}, {Risaliti}, {Fabbiano}, {Elvis},
  {Zezas}, \& {Karovska}}]{Wang_2010}
{Wang}, J., {Risaliti}, G., {Fabbiano}, G., {et~al.} 2010, \apj, 714, 1497

\end{thebibliography}
%

\end{document}